\documentclass[preprint]{aastex}
\usepackage{natbib}
\usepackage{graphicx}
 \usepackage{color}

\received{--}
\revised{--}
\accepted{--}
\slugcomment{ApJL 2010 in Press}
\shorttitle{The Genesis of Impulsive CMEs}
\shortauthors{Patsourakos, Vourlidas \& Stenborg}
\begin{document}

\title{The Genesis of an  Impulsive Coronal Mass Ejection observed at
  Ultra-High Cadence by AIA on \textit{SDO}} 

\author{S. Patsourakos\altaffilmark{1}, A. Vourlidas\altaffilmark{2},
  G. Stenborg\altaffilmark{3}}  

\altaffiltext{1}{University of Ioannina, Department of Physics,
  Section of Astrogeophysics, Ioannina, Greece} 

\altaffiltext{2}{Code 7663, Naval Research Laboratory, Washington, DC,
  USA}
\altaffiltext{3}{Interferometrics, Inc, Herdon, VA, USA}

%
%
\begin{abstract}
 The study of fast, eruptive events in the low solar corona is one of
  the science objectives of the Atmospheric Imaging Assembly (AIA)
  imagers on the recently launched \textit{Solar Dynamics Observatory
  (SDO)\/}, which take  full disk images in ten
  wavelengths with arcsecond resolution and 12 sec cadence.  
   We  study with AIA the 
  formation of an impulsive coronal mass ejection (CME) which occurred
  on June 13, 2010 and was associated with an M1.0 class
  flare. Specifically, we analyze the formation of the CME EUV bubble
  and its initial dynamics and
  thermal evolution in the low corona using AIA images in three wavelengths (171\AA, 193\AA, and
  211\AA). We derive the first ultra-high cadence measurements of the
  temporal evolution of the CME bubble aspect ratio (=bubble-height/bubble-radius). Our main result
  is that the CME formation undergoes three phases: it starts with a
  slow self-similar expansion followed by a fast but short-lived
($\sim$ 70 sec) period of strong
  lateral over-expansion  which essentially creates the CME. Then the CME
  undergoes another phase of self-similar expansion until it exits the
  AIA field of view. During the studied interval, the CME height-time profile
  shows a strong, short-lived, acceleration followed by
  deceleration. The lateral overexpansion phase coincides with the
  deceleration phase. The impulsive flare heating and CME acceleration
  are closely coupled. However, the lateral overexpansion of the CME
  occurs during the declining phase and is therefore not linked to
  flare reconnection.  In addition, the multi-thermal analysis of the
	    bubble does not show significant evidence of temperature change.

\end{abstract}

\keywords{Sun: coronal mass ejections (CMEs)}

\section{Introduction}
Synoptic observations of Coronal Mass Ejections (CMEs) have recorded
thousands of events over the last 15 years leading to a decent
understanding of their basic physical parameters
(e.g., \citet{2006SSRv..123..127S}). However, their initiation and driving
mechanisms remain unclear \citep{2006SSRv..123..251F} because their
formation and initial stages of evolution occur in the low corona
where, until recently, observational coverage was sparse and cadence
was low. 

Observationally, we can decompose the CME evolution into three stages: initiation,
acceleration, and propagation \citep{2006ApJ...649.1100Z}. Height-time
measurements of CME fronts, using EUV imagers and white light
coronagraphs, have shown a close correspondence between the initial
CME acceleration phase and the impulsive phase of the soft X-ray (SXR)
flare
\citep{2004ApJ...604..420Z,2008ApJ...673L..95T,2010ApJ...712.1410T}
and have established the CME acceleration profile as a proxy to the energy
release. Detailed acceleration profiles, especially for impulsive
CMEs, provide important constraints for all CME initiation models. 

The understanding of CME formation is lacking
behind the understanding of CME acceleration due to cadence and field of view
restrictions. Most CME models agree that the final
ejected structure is a magnetic fluxrope which may correspond
to the cavity observed in 3-part CMEs in the outer corona
\citep{prasad:07}.  But we do not yet know if the fluxrope
preexists, if it forms on-the-fly or if the final structure is a
combination of the above processes. The answer holds serious
implications for the relative importance of ideal versus non-ideal
mechanisms for the eruption process \citep[e.g.,][and references
therein]{2010ApJ...718.1388D}. Impulsive CMEs are of particular
interest because they tend to be associated both with strong flares
and particle acceleration. However, the formation of impulsive CMEs
is a particular challenge for existing instrumentation. They form low in the
corona amidst significant pre-existing fine scale structure, they interact
violently with their surroundings generating disturbances (such as
waves and loop deflections that can obscure the evolution of the
actual CME), and most of their energy release occurs within very short time
intervals, at or below the typical cadence of most EUV or white light
imagers. 

\citet{2010arXiv1008.1171P} (hereafter, PVK10) analyzed relatively
high cadence (75 - 150 sec) EUV STEREO observations of an impulsive
CME. The CME had the appearance of a bubble and it was
shown that the bubble was an early instance of the CME fluxrope seen
higher up in the coronagraph field of view. The wide separation
between the STEREO observations ($\sim 48^\circ$) allowed PVK10 to
perform detailed 3D modeling of the erupting bubble.  The modeling
showed that the bubble underwent a phase of lateral overexpansion
(i.e., the bubble radius increasing faster than its
height)
very
early in its evolution which coincided with the impulsive phase of the
flare.  The overexpansion transformed the bubble from a small,
arcminute-scale feature into a large CME-scale fluxrope structure
within 5 minutes. If this result holds in other events, it may explain
the discrepancy between the sizes of CMEs and post-eruptive
loops arcades and will clarify the true relationship between CMEs and
their source regions. To accurately identify, delineate and measure the
size of the erupting bubble, and the subsequent fluxrope, we need
high cadence, high sensitivity off-limb EUV images which only the EUV
imagers onboard the \textit{Solar Dynamics
  Observatory (SDO)\/} can provide. In this Letter, we present the first
multi-wavelength analysis of the early stages of a very impulsive CME
with ultra-high cadence (12 sec) and compare the 3D evolution of the
erupting bubble to the impulsive phase of the flare. Using three 
of the ten AIA channels sensitive to different temperatures we 
  search for evidence of plasma heating. The observations and data
analysis are presented in Section~2, and the results are discussed in
Section~3.

\section{Observations and Data Analysis}
We use EUV imaging observations from  AIA
\citep[AIA;][]{2006cosp...36.2600T} onboard the recently launched
SDO and the Extreme
Ultraviolet Imager \citep[EUVI;][]{2004SPIE.5171..111W} onboard the
\textit{Solar Terrestrial Relations Observatory (STEREO)\/} Ahead spacecraft
(STA). The eruption occurred on June 13, 2010 in active region NOAA
11079 which was located slightly behind the solar limb (S24W91). The
eruption was accompanied by a rather short duration M1.0 flare
starting at 05:33~UT, peaking at 05:39~UT and returning to pre-event
levels at around 06:43~UT. Associated with this eruption was a small
CME seen by the \textit{STEREO\/} coronagraphs. STA was situated $74^\circ$ west from
Earth, so AR11079 was $17^\circ$ east of STA's central
meridian. The region was therefore observed in quadrature by the
AIA and EUVI telescopes.

The eruption was observed as a limb event in all AIA channels with 12
sec cadence but here we use only the coronal channels centered at
171\AA, 193\AA\ and 211\AA \, (Level 1.5 images) and the EUVI-A 195
\AA\,channel (henceforth, we refer to a given channel simply by its
central wavelength). The EUVI-A 195 cadence was 2.5 min.

\begin{figure} 
\includegraphics[scale=0.8]{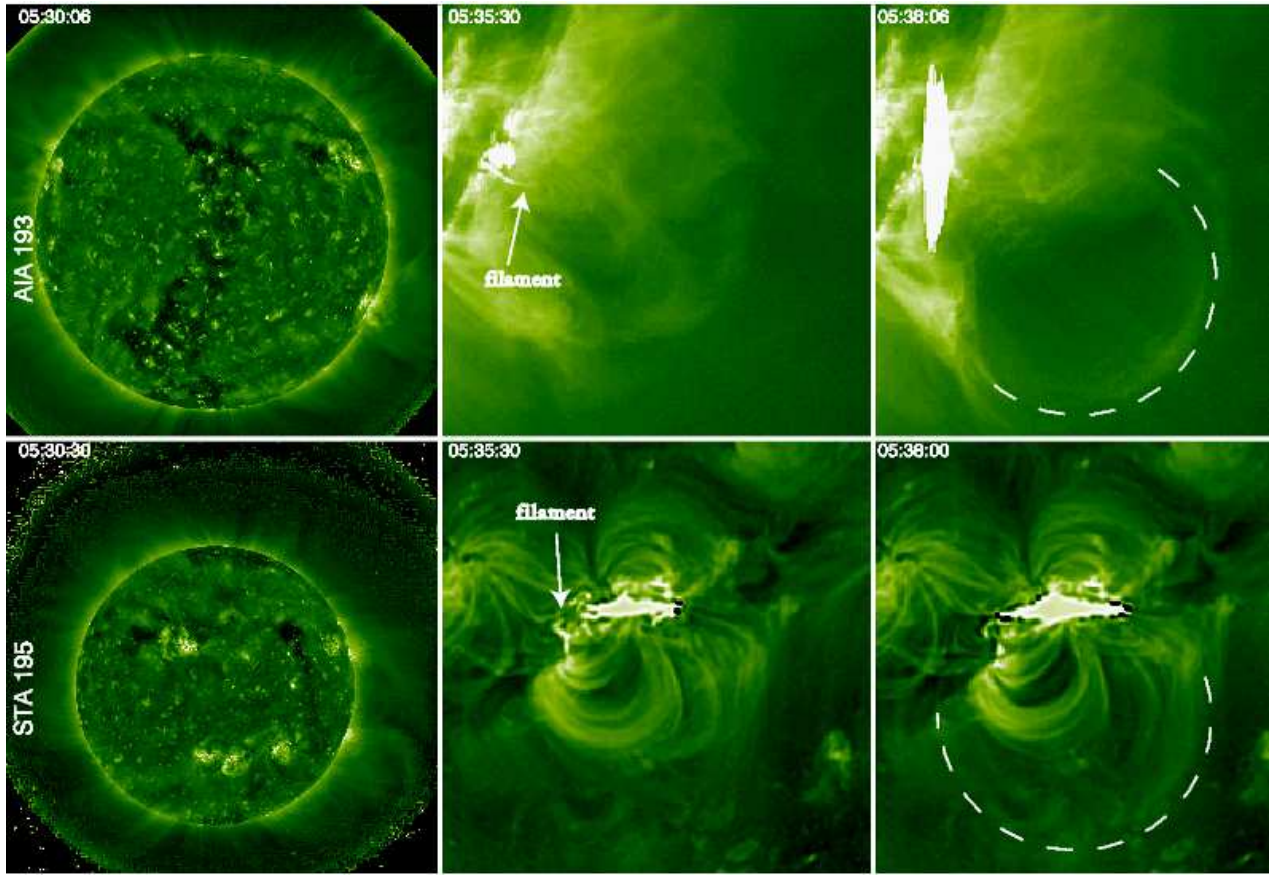}
  \caption{EUV Observations of the erupting active region in AIA 193
    (upper panel) and the EUVI-A 195 channel (lower panel). An
      untwisting filament apparently associated with the initiation
    of the eruption is marked by arrows. The dashed lines outline the
    expanding bubble. EUVI-A is situated $74^\circ$ West from
    Earth. }\label{fig:context}
\end{figure}

The active region exhibited considerable activity throughout the day
with loops rising and brightening but no major ejection was observed
before our event. We focus on the eruption, and concentrate on the
time interval from 05:30 to 05:43~UT. We provide two movies online in
AIA 193 and EUVI-A 195 (movie1.mpg, movie2.mpg). Snapshots from
these movies are shown in Figure~\ref{fig:context}. The image contrast
is enhanced with a wavelet technique \citep{2008ApJ...674.1201S}. Only
the large scale brightness component has been removed. 

The EUVI-A and AIA observations show the slow rise of loops over the
active region for several minutes before the CME. Starting at $\sim$
05:33~UT, brightenings appear very close to the surface at the inner
core of the region.  At the same time, a filament is clearly seen
rising and heating (Figure~\ref{fig:context}, middle column).  The filament
seems to rotate as it rises which may indicate that it is kinking or
writhing. The rise of the filament is associated with the rise of
surrounding loops which eventually form an EUV cavity resembling a
bubble around 05:36~UT.  Taking advantage of the two viewpoints
afforded by the EUVI-A and AIA observations we use triangulations to
locate the erupting filament and the bubble in the two telescopes (see
features marked in the righttmost column of
Figure~\ref{fig:context}).

\subsection{Formation of the bubble}

\begin{figure}
\includegraphics[scale=0.8]{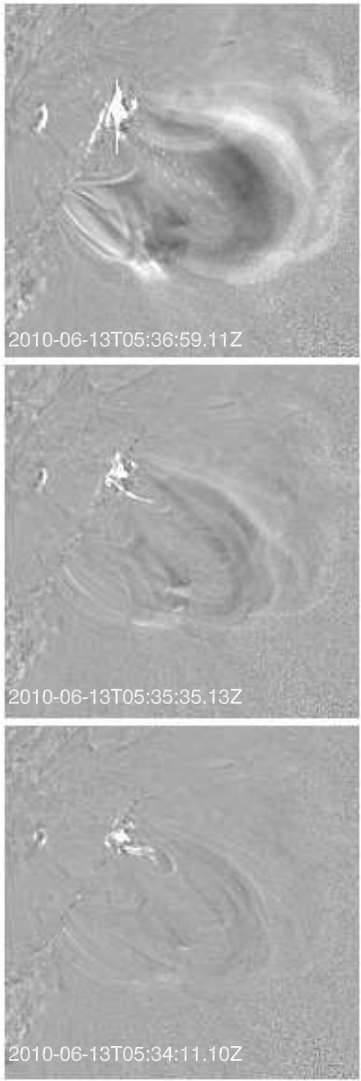}
  \caption{Representative AIA base-ratio 171 frames
{\it before} and {\it towards} the EUV bubble formation.
The ratio increases with from black to white within the range 0.3-1.4.} \label{fig:arcade2rope}
\end{figure}

Thanks to the high cadence of the AIA images, we can study the
formation and evolution of the EUV bubble in great detail. We provide
online a base-ratio movie in 171 (movie3.mpg), where we divide each
wavelet-enhanced image by a pre-event image at
05:30:23~UT. Representative snapshots from the movie are in
Figure~\ref{fig:arcade2rope}.  The motion of the rising filament
appears as twisting motion in AIA and straightening motion in
EUVI-A. This pattern seems consistent with writhe rather than kink
\citep{2010A&A...516A..49T}. As seen in AIA, the filament
rises, twists and then starts falling towards the surface (after
$\sim$ 05:33:11~UT). At that time, surrounding loops with seemingly
different orientations begin to slowly grow (e.g., alternating
black-white patterns in the lower two panels in
Figure~\ref{fig:arcade2rope}).  Simultaneously, a ribbon-like
intensity front propagates away from the filament along a
south-east direction (upper two frames,
Figure~\ref{fig:arcade2rope}). The extent of this front roughly matches
the lateral extension of the expanding loops. In the meantime, the
expanding loops seem to pile up at the surface of the forming bubble
(upper panel of Figure~\ref{fig:arcade2rope}). The bubble becomes
clear at $\sim$ 05:36~UT and continues to grow. After 2 minutes the
bubble begins a strong lateral expansion which we analyze
later. This time also signals the launch of an EUV
wave around the bubble (movie1.mpg) but the wave analysis
will be reported
elsewhere.  The outer rim of the bubble becomes progressively thinner
and dimmer after $\sim$05:38:23~UT. After $\sim$05:42~UT, it increasingly
becomes difficult to trace its upper part. The bubble exits the AIA field of view at $\sim$05:45~UT.

\subsection{Evolution of the bubble aspect ratio}

Following PVK10, we follow the evolution of the bubble
\textit{aspect ratio} to quantify the early 3D evolution of
the CME.  Unfortunately, the very impulsive nature of the event does
not permit detailed two-viewpoint fitting of the
bubble, as was done by PVK10. There is only one frame with the bubble
fully formed in both EUVI-A 195 and AIA 193 (third column,
Figure~\ref{fig:context}). 

\begin{figure} 
\includegraphics[scale=0.8]{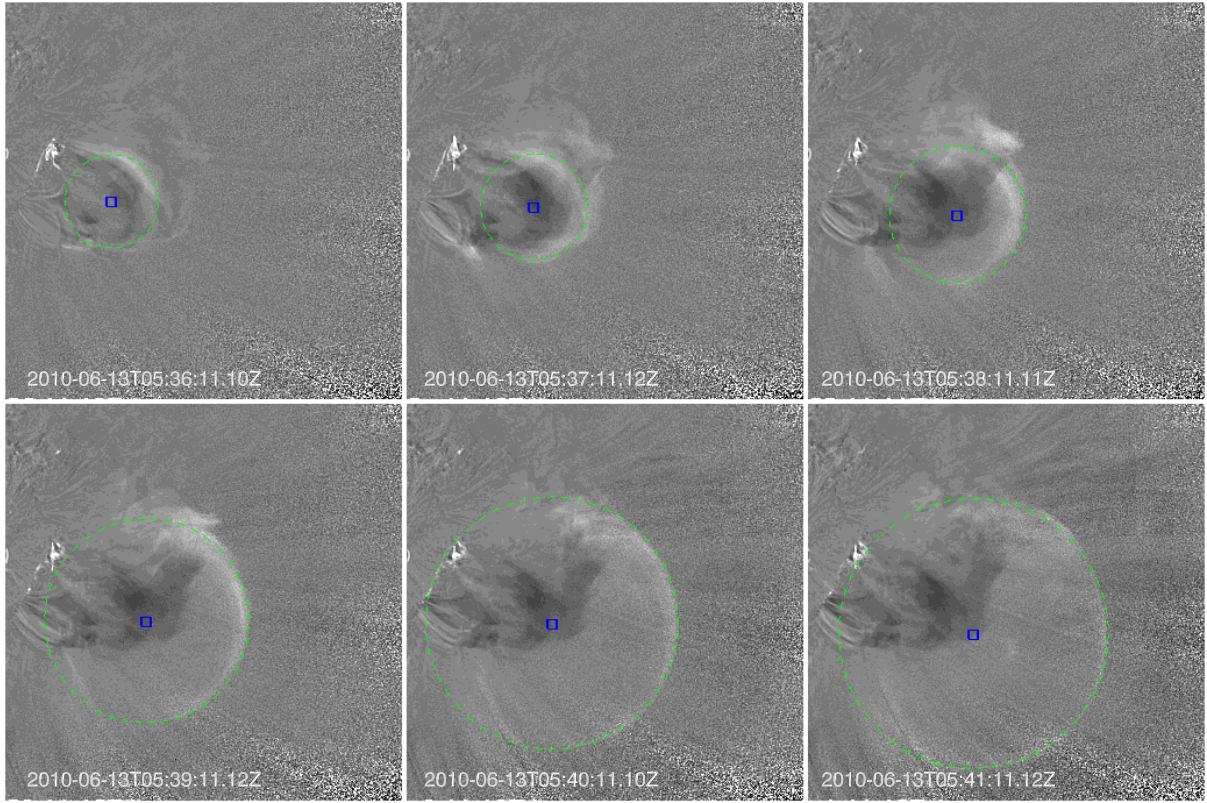}
\caption{Representative circular fits (green lines) of the bubble in AIA 171
  base-ratio images. The blue box marks the bubble center. The image at 05:30:23~UT is used as the base.} \label{fig:fit}
\end{figure}

We are constrained to use the AIA data only. Our approach is to fit
circles to the bubble rim using the 171 base-ratio images for the
interval 05:35:23-05:42:23~UT. There are 36 images in all.  We do not
use other times because it was difficult to identify a significant
fraction of the bubble rim with reasonable confidence due to
either its slow evolution or faint signal. For each 171 base-ratio
image, we manually select 10 points along the bubble rim and fit a
circle to these points.  We assume an error of 5 AIA pixels ($\sim 3$
arcsec) in the determination of each point. This rather conservative
estimate is the median of the half-width of the bubble rim for the
36 images. Figure~\ref{fig:fit} shows examples of the circle fits and demonstrates that a circle
represents a good approximation to the upper section of the bubble.
To further check the consistency of the circular assumption, we
perform a circular fit to the EUVI-A bubble at 05:38:00~UT which
yields a radius within 20\% of the AIA 193 fit at 05:38:06~UT
(Figure~\ref{fig:context}).  This agreement indicates that the
bubble can be reasonably approximated by a 3D sphere.

The circular fits supply the radius and center height, relative to the
solar surface, of the bubble. Hence the ratio between these two
quantities provides the {\it aspect ratio} of the erupting bubble (Figure~\ref{fig:mes}(a)).  The
error bars are determined by standard error propagation of the
uncertainties of the height and radius measurements. If we assume that
the bubble represents a rising fluxrope seen edge-on, we can interpret
the derived aspect ratio as the ratio of the major to the minor axis
of the erupting fluxrope. \footnote{The spherical approximation applies only to the upper section of the flux-rope
and not to its legs.}

The aspect ratio has a constant
value of $\approx$ 1.7 between $\sim$ 05:36:00~UT and $\sim$
05:38:20~UT. The ratio decreases rapidly, within approximatively 70 sec, 
to a value of $\approx 1$ and remains relatively constant thereafter. The "jump" in the aspect
ratio before $\sim$ 05:36~UT is possibly due to the
uncertainty in the bubble determination at that time.

\subsection{Bubble acceleration profile and flare dynamics}

\begin{figure}
\includegraphics[scale=1.0]{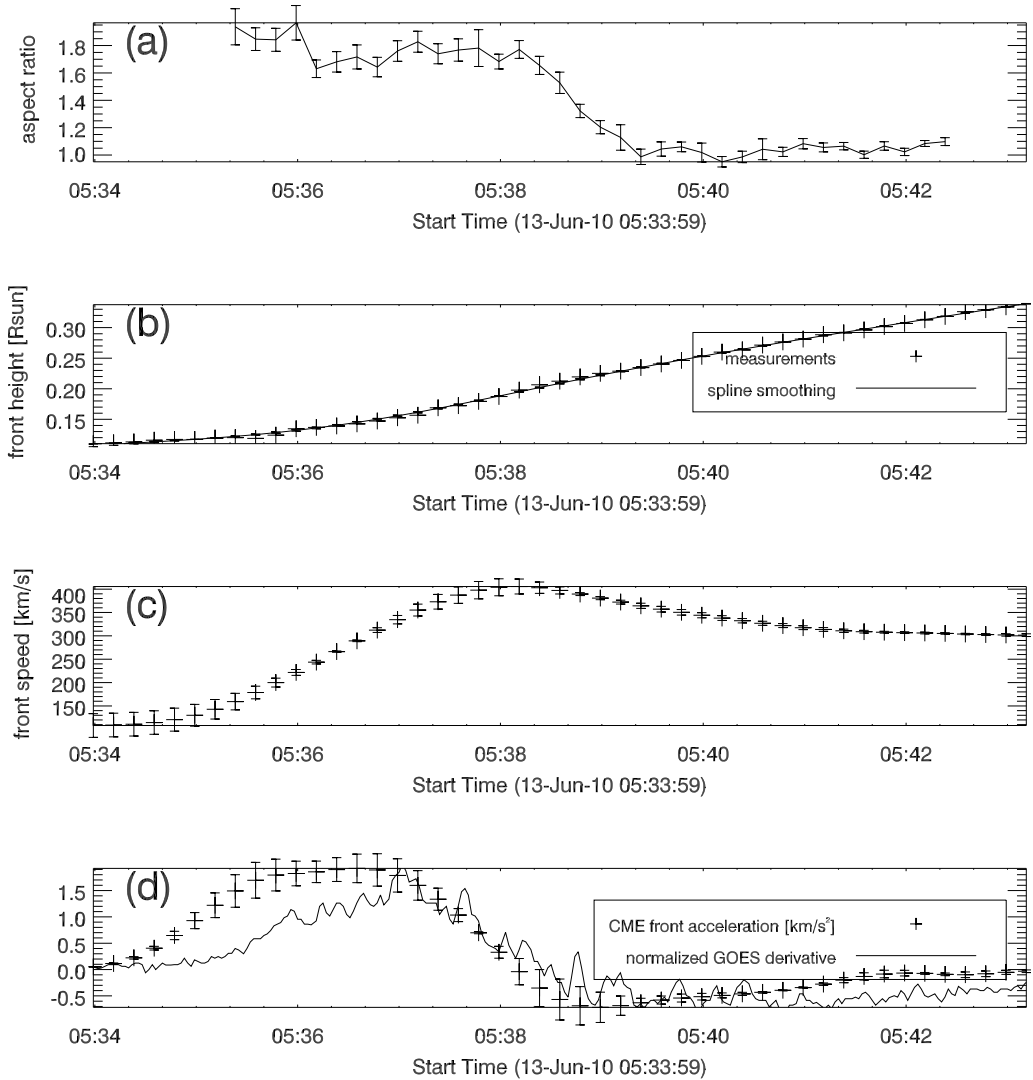}
  \caption{ (a) Temporal evolution of the aspect ratio of the bubble.
    (b) $HT$ measurements of the bubble front (crosses). The points
    are connected with spline smoothing curve (solid line). (c) Speed
    versus time curve prouced by differentiating the smoothed $HT$
    measurements.  (d) Acceleration versus time curve (crosses)
    over-plotted on the time-derivative of the GOES SXR curve (solid
    line).
  } \label{fig:mes}
\end{figure}

Following common practice, we measure the height-time ($HT$) evolution
of the EUV bubble in 171 by tracking a point in the bubble front
between 05:34:00-05:43:11~UT.  The resulting 47 measurements are shown
in Figure \ref{fig:mes}(b).  As done frequently, the $HT$ data are
first smoothed to reduce small-scale fluctuations
\citep[e.g.,][]{2007SoPh..241...85V}.  We use a smoothing cubic
  spline scheme which minimizes a function consisting of the sum of a
  ${\chi}^{2}$ fit of the data with a cubic spline plus a penalty
  function proportional to the second derivative of the cubic spline
  multiplied by a user-supplied factor $0 \leq spar \leq 1$,e.g.
  \citet{weisberg_2005}.  Smoothing factors between $0.7-0.8$ provide
the best compromise between noisy (zero or small $ spar$) and
extremely smoothed (large $spar$) acceleration profiles. We adopt the
median of the above interval ($spar=0.75$) for the smoothing of the
$HT$ data (Figure~\ref{fig:mes}(b)) which are then differentiated with
respect to time to obtain velocity and acceleration profiles (panels
(c) and (d) of Figure~\ref{fig:mes}).  The error bars are derived from
the maximum residuals between the curves calculated with the median
$spar$ and those for $spar$ 0.7 and 0.8. The errors in the speed
  and acceleration profiles resulting from the use of 5-pixel error
  estimate in the HT measurements were very small and were thus
  neglected.  The results show that the bubble accelerates rapidly
(within 4 min) to a speed of $\approx$ 400 $\mathrm{km\,{s}^{-1}}$ and
then starts to decelerate (Figure~\ref{fig:mes}c). The acceleration
profile is characterized by a short-lived ($FWHM$ $\approx$2 min)
acceleration pulse followed by deceleration. The latter is expected to
occur even in the inner corona, for impulsively accelerated CMEs
(e.g., \citet{2010ApJ...712.1410T}). This may be due to solar
  wind drag as the CME speed exceeds the local solar wind speed. The
deceleration is verified by the LASCO C2 measurements which show a speed of $\approx$ 200 $\mathrm{km\,{s}^{-1}}$.

The acceleration profile corresponds closely to the time derivative of
the GOES soft X-ray (SXR) light curve, a proxy for hard X-rays (HXR)
and hence for the flare reconnection\footnote{RHESSI was observing the
  Crab Nebula during our event.}, (Figure~\ref{fig:mes}(d)), as
generally expected
\citep[e.g.,][]{2004ApJ...604..420Z,2008ApJ...673L..95T,2010ApJ...712.1410T}.
The slower SXR rise profile arises from the occultation of a
significant part of the flaring region from Earth. The SXR loops
contribute to the GOES lightcurve only after they rise above the solar
limb.  The SXR levels are high when the bubble forms ($\sim$05:36~UT).
Both the SXR derivative (and hence reconnection) and the bubble
acceleration are close to zero when the strong lateral expansion
starts. It is an indication that the expansion may be driven by mostly
an {\it ideal} process. We will return to this important point in the
Discussion.

\subsection{Multi-temperature behavior of the bubble}
We investigate the thermal behavior of the expanding bubble
using the almost simultaneous AIA observations with filters sensitive
to coronal plasmas of different temperatures. We use the
observations in 171, 193, and 211 channels with peak responses at 0.8,
1.25, and 1.6 MK, respectively.

\begin{figure}
\includegraphics[scale=0.8]{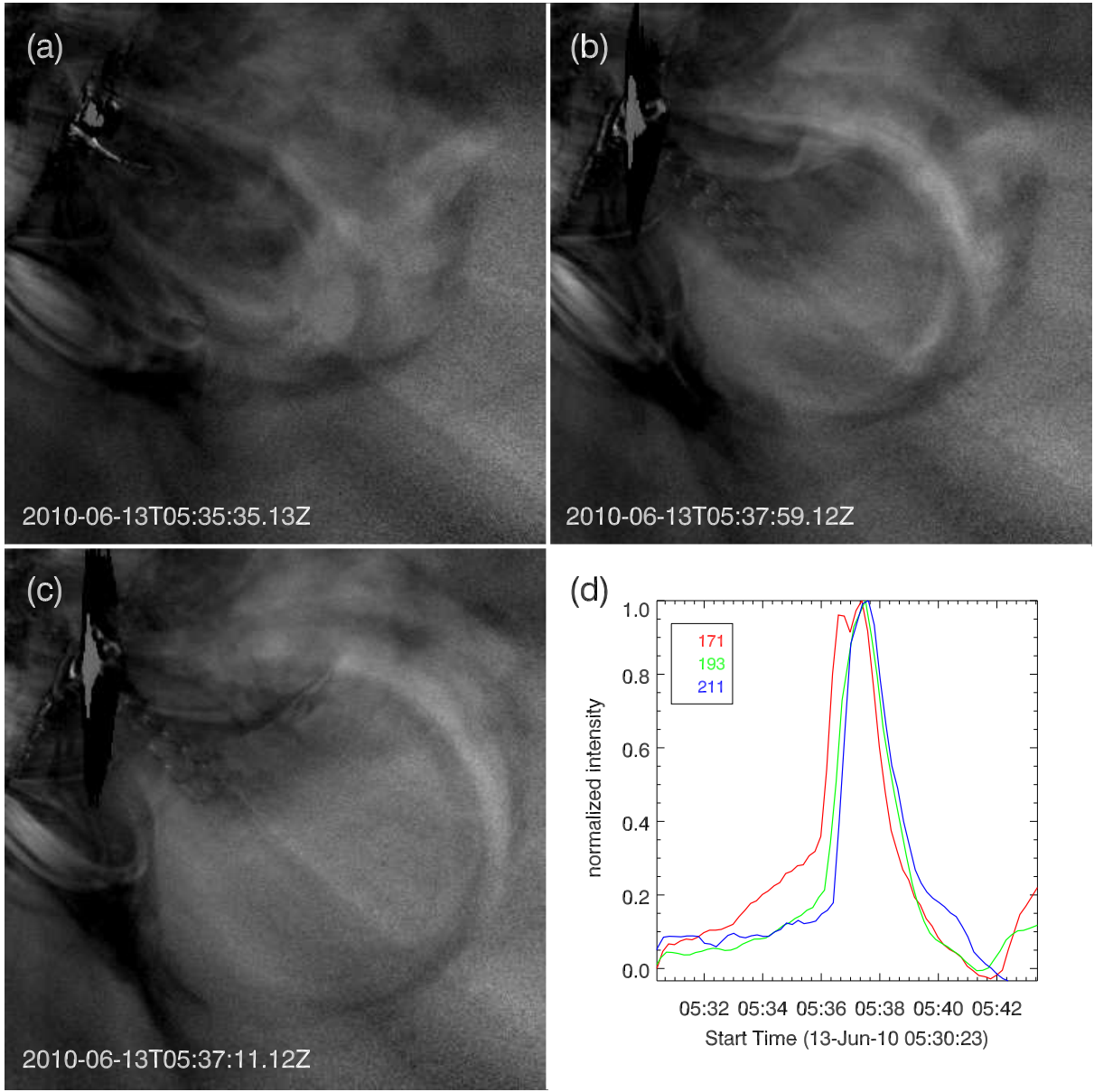}
  \caption{(a)-(c) AIA 171/193 intensity ratio snapshots.
(d) Light-curves across the bubble rim in 171, 193, and 211 with
pre-event intensity levels subtracted. The times on the images
correspond to 171 and the 193 images trail by 5 sec.} \label{fig:multi}
\end{figure}

We posted an online movie of the $171/193$ intensity ratio
during the event (movie4.mpg). The observations between the two channels are
separated by only 5 sec which corresponds to a displacement of around
4 pixels at the maximum speed of $\approx300\,
\mathrm{km\,{s}^{-1}}$. This separation is smaller than the 10 pixel
full width of the erupting rim and so it has minimal impact in the
calculated intensity ratios. Snapshots from the intensity ratio movie
are shown in Figure~\ref{fig:multi}.  Brighter (darker) areas could
signify relatively larger (lower) amounts of warmer (hotter) plasmas
\footnote{Note that filter-ratios are not
  strictly monotonic functions of temperature over large temperature
intervals.}.  Such areas seem to
remain unaltered which implies that the
temperature does not vary substantially during the event. In Figure
\ref{fig:multi}, the bright/dark segments of the bubble
remain as such during its evolution. 

To further quantify the thermal evolution of the bubble, we construct
lightcurves by extracting the intensity along the propagation path
between 05:30~UT and 05:42~UT in each channel\footnote{The intensities
  in the hotter channels, i.e. 335, 94, 131 were affected by
  diffraction patterns from the flare emission and were unsuitable for
  quantitative analysis.}.  The intensities at 5:30~UT are subtracted from each exposure. The peak intensity for the
base-subtracted lightcurves is used as a proxy for the bubble
intensity.  All three lightcurves peak nearly simultaneously and at about the
same time as the
strong lateral expansion of the bubble (Figure~\ref{fig:mes}).  Also,
the lightcurves have similar lifetimes to the acceleration pulse
($FWHM$ $\approx$ 2 min). The 171 lightcurve, however,
increases at a slower rate than either the 193 or 211 curves.  This
may be because the loops rising from the AR core were
emitting mostly in 171. The hotter emissions seem to
lie around these 171 loops and the bubble in those channels seems to
form as a result of the initial expansion of the 171 structures. The
discussion above suggests that the bubble temperature did not
vary substantialy during its expansion.

\section{Discussion and Conclusions}

Arguably, our most interesting finding is the strong lateral
overexpansion of the bubble that in essence 'inflates' an initially
small-scale feature into a CME-scale structure.  This period of
inflation starts {\it after} the main phase of flare heating and CME
acceleration and argues strongly for {\it ideal MHD} rather than a
reconnection-related process.  The observed overexpansion is a
considerable challenge for any CME model. The overexpansion may be
triggered when the expanding CME 'exits' above the AR loop core and
encounters weaker magnetic fields. To achieve pressure balance with its
surroundings, the CME bubble must expand laterally
\citep{2007ApJ...668.1221M}.  Another explanation may be flux
conservation around a rising fluxrope of decreasing current (PVK10).
However, the overexpansion in PVK10 was relatively
well-synchronized with the impulsive flare phase and hence a
reconnection-based driver was possible.  The differences between these impulsive CME events indicate that the characteristics of lateral
overexpansion in the early phase of CMEs vary from event to event
making their study a potentially sensitive diagnostic of CME evolution
and coronal properties. For instance, a deeper study of this event may
  supply estimates on the magnetic field strength of the
  regions in the vicinity of the eruption site. A detailed MHD
  modeling of the PKV10 event to determine the relative role of ideal
  vs non-ideal processes is underway.

Our multi-temperature observations show no appreciable temperature
changes in the bubble during the event. If significant plasma heating
(or cooling) was taking place, it would have led to temporally
displaced lightcurves in the different channels
\citep[e.g.,][]{robbrecht_wang_2010}.  This does not necessarily
preclude heating (and then cooling) of the bubble plasma during its
formation. According to standard flare-CME theory, field lines
reconnecting in the vertical current sheet underneath the erupting
flux become part of the CME bubble
\citep[e.g.,][]{lin_ray_van_2004}. If these lines fill with hot plasma
(i.e., by evaporation) at a slower rate than the eruption, then no
temperature changes may be detected.  Indeed, for the analyzed AIA
channels ($\approx$ 0.5-3 MK), the evaporation speeds are around 100
$\mathrm{km\,{s}^{-1}}$ \cite[e.g.,][]{2009ApJ...699..968M}---slower
than the CME speed($\sim 200-400\, \mathrm{km\,{s}^{-1}}$). More analysis is needed to test these ideas further.  

While reconnection may not be important in the lateral overexpansion
of the bubble, it could be a {\it significant\/} factor for its
formation. The bubble forms from a set of pre-existing loops at
varying orientations during the main flare phase. This is consistent
with the transformation of a loop arcade to a fluxrope structure and
is predicted by several CME models
\citep[e.g.,][]{2008ApJ...683.1192L}. However, the possible eruption
trigger lies in the rise and possible instability of a small
filament. Such structure could be related to a pre-existing
fluxrope. At any rate, our observations of the initially very small
bubble radius ($\sim 0.03 R_{\odot}$) set an upper limit for the size
of any pre-existing flux rope. We suggest that the observed lateral
overexpansion may well the process through which eruptions starting
small in the corona become large-scale CMEs further out. Analysis of
more events is required to establish or refute this suggestion.

To summarize, we have presented EUV observations of the genesis of an
impulsive CME. By taking full advantage of the unique SDO AIA
capabilities ---its unprecedented high cadence (12 sec) and
multi-temperature coverage--- we were able to resolve the various
stages of the event in great detail.  Our main findings are:
\begin{itemize}
\item A set of slowly rising loops, possibly triggered by a rising and
  maybe kinking or writhing filament, is transformed into an EUV
  bubble within 2 minutes.
\item The EUV bubble forms when both flare heating and CME
  acceleration are at their maximum levels.
\item The bubble experiences a 70 sec period of strong lateral
  overexpansion followed by self-similar evolution.
\item The lateral over-expansion starts when flare reconnection and
  CME acceleration are well through their peaks.
\item The bubble rim emission shows no significant evidence of
  temperature change.
\end{itemize}

\begin{acknowledgements}
  The AIA data used here are courtesy of \textit{SDO\/} (NASA) and the
  AIA consortium.  We thank the AIA team for the easy access to
  calibrated data and the referee for useful comments.  The SECCHI data are courtesy of \textit{STEREO\/}
  and the SECCHI consortium.

\end{acknowledgements}


\begin{thebibliography}{20}
\expandafter\ifx\csname natexlab\endcsname\relax\def\natexlab#1{#1}\fi

\bibitem[{{D{\'e}moulin} \& {Aulanier}(2010)}]{2010ApJ...718.1388D}
{D{\'e}moulin}, P., \& {Aulanier}, G. 2010, \apj, 718, 1388

\bibitem[{{Forbes} {et~al.}(2006){Forbes}, {Linker}, {Chen}, {Cid}, {K{\'o}ta},
  {Lee}, {Mann}, {Miki{\'c}}, {Potgieter}, {Schmidt}, {Siscoe}, {Vainio},
  {Antiochos}, \& {Riley}}]{2006SSRv..123..251F}
{Forbes}, T.~G., {et~al.} 2006, \ssr, 123, 251

\bibitem[{Lin {et~al.}(2004)Lin, Raymond, \& van
  Ballegooijen}]{lin_ray_van_2004}
Lin, J., Raymond, J.~C., \& van Ballegooijen, A.~A. 2004, The Astrophysical
  Journal, 602, 422–435

\bibitem[{{Lynch} {et~al.}(2008){Lynch}, {Antiochos}, {DeVore}, {Luhmann}, \&
  {Zurbuchen}}]{2008ApJ...683.1192L}
{Lynch}, B.~J., {Antiochos}, S.~K., {DeVore}, C.~R., {Luhmann}, J.~G., \&
  {Zurbuchen}, T.~H. 2008, \apj, 683, 1192

\bibitem[{{Milligan} \& {Dennis}(2009)}]{2009ApJ...699..968M}
{Milligan}, R.~O., \& {Dennis}, B.~R. 2009, \apj, 699, 968

\bibitem[{{Moore} {et~al.}(2007){Moore}, {Sterling}, \&
  {Suess}}]{2007ApJ...668.1221M}
{Moore}, R.~L., {Sterling}, A.~C., \& {Suess}, S.~T. 2007, \apj, 668, 1221

\bibitem[{{Patsourakos} {et~al.}(2010){Patsourakos}, {Vourlidas}, \&
  {Kliem}}]{2010arXiv1008.1171P}
{Patsourakos}, S., {Vourlidas}, A., \& {Kliem}, B. 2010, ArXiv e-prints

\bibitem[{Robbrecht \& Wang(2010)}]{robbrecht_wang_2010}
Robbrecht, E., \& Wang, Y.-M. 2010, The Astrophysical Journal, 720, L88

\bibitem[{{Schwenn} {et~al.}(2006){Schwenn}, {Raymond}, {Alexander},
  {Ciaravella}, {Gopalswamy}, {Howard}, {Hudson}, {Kaufmann}, {Klassen},
  {Maia}, {Munoz-Martinez}, {Pick}, {Reiner}, {Srivastava}, {Tripathi},
  {Vourlidas}, {Wang}, \& {Zhang}}]{2006SSRv..123..127S}
{Schwenn}, R., {et~al.} 2006, Space Science Reviews, 123, 127

\bibitem[{{Stenborg} {et~al.}(2008){Stenborg}, {Vourlidas}, \&
  {Howard}}]{2008ApJ...674.1201S}
{Stenborg}, G., {Vourlidas}, A., \& {Howard}, R.~A. 2008, \apj, 674, 1201

\bibitem[{{Subramanian} \& {Vourlidas}(2007)}]{prasad:07}
{Subramanian}, P., \& {Vourlidas}, A. 2007, \aap, 467, 685

\bibitem[{{Temmer} {et~al.}(2010){Temmer}, {Veronig}, {Kontar}, {Krucker}, \&
  {Vr{\v s}nak}}]{2010ApJ...712.1410T}
{Temmer}, M., {Veronig}, A.~M., {Kontar}, E.~P., {Krucker}, S., \& {Vr{\v
  s}nak}, B. 2010, \apj, 712, 1410

\bibitem[{{Temmer} {et~al.}(2008){Temmer}, {Veronig}, {Vr{\v s}nak},
  {Ryb{\'a}k}, {G{\"o}m{\"o}ry}, {Stoiser}, \& {Mari{\v
  c}i{\'c}}}]{2008ApJ...673L..95T}
{Temmer}, M., {Veronig}, A.~M., {Vr{\v s}nak}, B., {Ryb{\'a}k}, J.,
  {G{\"o}m{\"o}ry}, P., {Stoiser}, S., \& {Mari{\v c}i{\'c}}, D. 2008, \apjl,
  673, L95

\bibitem[{{Title} {et~al.}(2006){Title}, {Hoeksema}, {Schrijver}, \& {The Aia
  Team}}]{2006cosp...36.2600T}
{Title}, A.~M., {Hoeksema}, J.~T., {Schrijver}, C.~J., \& {The Aia Team}. 2006,
  in COSPAR, Plenary Meeting, Vol.~36, 36th COSPAR Scientific Assembly, 2600--+

\bibitem[{{T{\"o}r{\"o}k} {et~al.}(2010){T{\"o}r{\"o}k}, {Berger}, \&
  {Kliem}}]{2010A&A...516A..49T}
{T{\"o}r{\"o}k}, T., {Berger}, M.~A., \& {Kliem}, B. 2010, \aap, 516, A49+

\bibitem[{{Vr{\v s}nak} {et~al.}(2007){Vr{\v s}nak}, {Mari{\v c}i{\'c}},
  {Stanger}, {Veronig}, {Temmer}, \& {Ro{\v s}a}}]{2007SoPh..241...85V}
{Vr{\v s}nak}, B., {Mari{\v c}i{\'c}}, D., {Stanger}, A.~L., {Veronig}, A.~M.,
  {Temmer}, M., \& {Ro{\v s}a}, D. 2007, \solphys, 241, 85

\bibitem[{Weisberg(2005)}]{weisberg_2005}
Weisberg, S. 2005, Applied linear regression (John Wiley and Sons)

\bibitem[{{Wuelser} {et~al.}(2004){Wuelser}, {Lemen}, {Tarbell}, {Wolfson},
  {Cannon}, {Carpenter}, {Duncan}, {Gradwohl}, {Meyer}, {Moore}, {Navarro},
  {Pearson}, {Rossi}, {Springer}, {Howard}, {Moses}, {Newmark},
  {Delaboudiniere}, {Artzner}, {Auchere}, {Bougnet}, {Bouyries}, {Bridou},
  {Clotaire}, {Colas}, {Delmotte}, {Jerome}, {Lamare}, {Mercier}, {Mullot},
  {Ravet}, {Song}, {Bothmer}, \& {Deutsch}}]{2004SPIE.5171..111W}
{Wuelser}, J., {et~al.} 2004, in Society of Photo-Optical Instrumentation
  Engineers (SPIE) Conference Series, Vol. 5171, Society of Photo-Optical
  Instrumentation Engineers (SPIE) Conference Series, ed. {S.~Fineschi \&
  M.~A.~Gummin}, 111--122

\bibitem[{{Zhang} \& {Dere}(2006)}]{2006ApJ...649.1100Z}
{Zhang}, J., \& {Dere}, K.~P. 2006, \apj, 649, 1100

\bibitem[{{Zhang} {et~al.}(2004){Zhang}, {Dere}, {Howard}, \&
  {Vourlidas}}]{2004ApJ...604..420Z}
{Zhang}, J., {Dere}, K.~P., {Howard}, R.~A., \& {Vourlidas}, A. 2004, \apj,
  604, 420

\end{thebibliography}


\clearpage

\end{document}